\documentclass{article}

\usepackage{microtype}
\usepackage{graphicx}
\usepackage[T1]{fontenc}
\usepackage{subfigure}
\usepackage{booktabs} % for professional tables
\usepackage{svg}
\usepackage{multirow} % Add the multirow package

\usepackage{hyperref}

\usepackage[accepted]{icml2024}

\usepackage{amsmath}
\usepackage{amssymb}
\usepackage{mathtools}
\usepackage{amsthm}

\usepackage[capitalize,noabbrev]{cleveref}

\theoremstyle{plain}

\theoremstyle{definition}

\theoremstyle{remark}

\usepackage[textsize=tiny]{todonotes}

\makeatletter
\newcommand\footnoteref[1]{\protected@xdef\@thefnmark{\ref{#1}}\@footnotemark}
\makeatother

\icmltitlerunning{Fast Timing-Conditioned Latent Audio Diffusion}

\begin{document}

\twocolumn[
\icmltitle{Fast Timing-Conditioned Latent Audio Diffusion}



\icmlsetsymbol{equal}{*}

\begin{icmlauthorlist}
\icmlauthor{Zach Evans}{yyy}
\icmlauthor{CJ Carr}{yyy}
\icmlauthor{Josiah Taylor}{yyy}
\icmlauthor{Scott H. Hawley}{comp}
\icmlauthor{Jordi Pons}{yyy}
\end{icmlauthorlist}

\icmlaffiliation{yyy}{Stability AI.}
\icmlaffiliation{comp}{Belmont University, work done while at Stability AI}

\icmlcorrespondingauthor{Zach Evans}{zach@stability.ai}

\icmlkeywords{Machine Learning, ICML, Stable Audio, fast, timing-conditioned, latent audio diffusion,}

\vskip 0.3in
]



\printAffiliationsAndNotice{} % otherwise use the standard text.

\begin{abstract}

Generating long-form 44.1kHz stereo audio from text prompts can be computationally demanding. Further, most previous works do not tackle that music and sound effects naturally vary in their duration. 
Our research focuses on the efficient generation of long-form, variable-length stereo music and sounds at 44.1kHz using text prompts with a generative model.
Stable Audio is based on latent diffusion, with its latent defined by a fully-convolutional variational autoencoder. It is conditioned on text prompts as well as timing embeddings, allowing for fine control over both the content and length of the generated music and sounds.
Stable Audio is capable of rendering stereo signals of up to 95 sec at 44.1kHz in 8 sec on an A100 GPU. 
Despite its compute efficiency and fast inference, it is one of the best in two public text-to-music and -audio benchmarks and, differently from state-of-the-art models, can generate music with structure and stereo sounds.

\end{abstract}

\section{Introduction}
\label{introduction}

The introduction of diffusion-based generative models \cite{sohldickstein2015deep, ho2020denoising} has 
lead to rapid improvements in the quality and controllability of generated images \cite{podell2023sdxl}, video \cite{blattmann2023align}, and audio \cite{rouard2021crash,liu2023audioldm}.

One challenge is that diffusion models working within the raw signal space tend to be computationally demanding during both training and inference. Diffusion models working in the latent space of a pre-trained autoencoder, termed “latent diffusion models” \cite{rombach2022highresolution}, are significantly more compute efficient. 
Working with a heavily downsampled latent representation of audio allows for much faster inference times compared to raw audio, and also allows generating long-form audio (e.g., 95 sec).

Another challenge with audio diffusion models is that those are usually trained to generate fixed-size outputs \cite{huang2023noise2music}, e.g., a model trained on 30 sec audio chunks will generate 30 sec outputs. This is an issue when training on and trying to generate audio of varying lengths, as is the case when generating full songs or sound effects.
Hence audio diffusion models are commonly  trained on randomly cropped chunks from longer audios, cropped or padded to fit the diffusion model’s training length. With music, e.g., this causes the model to generate arbitrary sections of a song, which may start or end in the middle of a musical phrase.

Stable Audio is based on a latent diffusion model for audio conditioned on a text prompt as well as timing embeddings, allowing for control over the content and length of the generated music and sound effects. This additional timing conditioning allows us to generate audio of a specified (variable) length up to the training window length. 
Due to the compute efficient nature of latent diffusion modeling, it can generate long-form content in short inference times.
It can render up to 95 sec (our training window length) of stereo audio at 44.1kHz in 8 sec on an A100 GPU (40GB VRAM).

The commonly used metrics for generative audio are designed to evaluate short-form mono signals at 16kHz~\cite{kilgour2018fr}. Yet, our work focuses on generating long-form full-band stereo signals. We propose: (i)~a Fréchet Distance based on OpenL3 embeddings \cite{cramer2019look} to evaluate the plausibility of the generated long-form full-band stereo signals, (ii)~a Kullback-Leibler divergence to evaluate the semantic correspondence between lengthy generated and reference audios up to 32kHz, and (iii)~a CLAP score to evaluate how long-form full-band stereo audios adhere to the given text prompt. 
We also conduct a qualitative study, assessing audio quality and text alignment, while also pioneering the assessment of musicality, stereo correctness, and musical structure. 
We show that Stable Audio can obtain state-of-the-art results on long-form full-band stereo music and sound effects generation from text and timing inputs. 
We also show that, differently from previous works, Stable Audio is also capable to generate structured music (with intro, development, outro) and stereo sound effects.

Code to reproduce our model/metrics and demos is online\footnote{Model: \href{https://github.com/Stability-AI/stable-audio-tools}{https://github.com/Stability-AI/stable-audio-tools}.\linebreak Metrics: \href{https://github.com/Stability-AI/stable-audio-metrics}{https://github.com/Stability-AI/stable-audio-metrics}.\linebreak  Demo: \hspace{5.7mm}\href{https://stability-ai.github.io/stable-audio-demo}{https://stability-ai.github.io/stable-audio-demo}.}.

\section{Related Work}
\label{relatedwork}

\textbf{Autoregressive models} --- WaveNet \cite{oord2016wavenet} autoregressively models quantized audio samples, but is slow during inference because it operates with waveforms. 
Recent autoregressive models addressed this by operating on a quantized latent space, enabling faster processing. Jukebox \cite{dhariwal2020jukebox} relies on a multi-scale approach to encode music into a sequence of quantized latents and subsequently models them using autoregressive transformers.  
Recent work such as MusicLM \cite{agostinelli2023musiclm} and MusicGen \cite{copet2023simple} utilize a similar approach and also autoregressively model quantized latent sequences.
However, unlike Jukebox, such models are conditioned on text prompts rather than on artist, genre, and/or lyrics. 
Autoregressive models similar to MusicLM (AudioLM) and MusicGen (AudioGen) have also been used for sound synthesis \cite{borsos2023audiolm,kreuk2022audiogen} and for generating music accompaniments from singing \cite{donahue2023singsong}.
Our work is not based on autoregressive modeling.

\textbf{Non-autoregressive models} --- Parallel WaveNet \cite{oord2018parallel} and adversarial audio synthesis \cite{donahue2018adversarial,pasini2022musika} were developed to tackle the computational inefficiencies inherent in autoregressive modeling. 
Recent works like VampNet \cite{garcia2023vampnet}, %SoundStorm \cite{borsos2023soundstorm}
StemGen \cite{stemgen} and MAGNeT \cite{ziv2024masked} are based on masked token modeling \cite{chang2022maskgit}. These are for creating musical variations, 
generating additional stems for a given song, and to efficiently synthesize music and sounds, respectively.
Flow-matching generative modeling \cite{vyas2023audiobox} was also recently introduced for speech and sounds synthesis.
Our work is not based on any of the non-autoregressive models above.

\textbf{End-to-end diffusion models} ---  CRASH \cite{rouard2021crash} was proposed for unconditional drums synthesis,
DAG \cite{pascual2023full} for class-conditional sounds synthesis, Noise2Music \cite{huang2023noise2music} for text-conditional music synthesis, and \citet{mariani2023multi} built an end-to-end diffusion model capable of both music synthesis and source separation.
Our work is also based on diffusion, albeit not in an end-to-end fashion. Rather, it involves latent diffusion due to its computational efficiency.

\textbf{Spectrogram diffusion models} --- Riffusion \cite{Forsgren_Martiros_2022} fine-tuned Stable Diffusion to generate spectrograms from text prompts, 
\citet{hawthorne2022multi} addressed MIDI-to-spectrogram generation, and  
CQT-Diff \cite{moliner2023solving} relied on CQT spectrograms for bandwidth extension, inpatining, and declipping.
An additional step is required to render waveforms from magnitude spectrograms.
Our work is also based on diffusion, albeit it does not rely on spectrogram-based synthesis.

\break

\textbf{Latent diffusion models} ---  Mo\^{u}sai \cite{schneider2023mousai} and AudioLDM \cite{liu2023audioldm} pioneered using latent diffusion for text-to-music and -audio. 
Their main difference being that Mo\^{u}sai decodes latents onto waveforms through a diffusion decoder, while AudioLDM decodes latents onto spectrograms which are then inverted to waveforms with HiFi-GAN \cite{kong2020hifi}.
AudioLDM2 \cite{liu2023audioldm2} extends AudioLDM to also synthesize speech by using a shared representation for music, audio, and speech to condition the latent diffusion model.
JEN-1 \cite{li2023jen1} is an {\textit{omnidirectional}} latent diffusion model trained in a multitask fashion. 
JEN-1 Composer \cite{yao2023jen} is its extension for multi-track music generation.
\citet{apple} explored sampling-time guidance for both end-to-end and latent diffusion models. 
All previous works constrain the latent to be normalized, often with a variational autoencoder (VAE). The exceptions being JEN-1, which runs over a dimensionality reduced latent that is normalized based on the mean and covariance, and Mo\^{u}sai that simply uses a tanh.
Our work is also based on latent diffusion, and we normalize latents by using a VAE. Appendix \ref{appendix:related} includes further discussion on related latent diffusion models.

\textbf{High sampling rate and stereo generation} ---  
Mo\^{u}sai and JEN-1 generate 48kHz stereo music. 
AudioLDM2 can generate 48kHz mono music.
\citet{apple} generates 44.1kHz stereo music.
No other prior works generate music up to the standard specifications of commercial music (44.1kHz stereo). DAG and AudioLDM2 generate 48kHz mono sounds, and we are not aware of prior works tackling stereo sound synthesis.
Our work focuses on generating 44.1kHz stereo music and sounds from text prompts.

\textbf{Text embeddings} --- CLAP \cite{wu2023large} and T5-like \cite{raffel2020exploring, ghosal2023text} text embeddings are commonly used because of their open-source nature. CLAP relies on a contrastive (multimodal) language-audio pretraining, and T5 is a large language model. Further, MusicLM uses MuLan \cite{huang2022mulan}, that is also based on contrastive language-audio pretraining but on their private dataset. 
Our work relies on a CLAP-based model trained in a contrastive language-audio fashion on our dataset.

\textbf{Fast generation of variable-length, long-form audio} --- Autoregressive models can generate long-form audio of variable length due to their sequential (one-sample-at-a-time generation) nature, but are slow at inference time.
Previous non-autoregressive models were trained to generate up to 20 sec long music \cite{stemgen}.
Previous end-to-end and latent diffusion models were trained to generate up to 30 sec long music \cite{huang2023noise2music,apple,lam2024efficient}, with the exception of Mo\^{u}sai that was trained to generate 44 sec. %\footnote{Although Mo\^{u}sai demos show generations of 87 sec. We could not find details about how those demos were generated.}
Hence, previous works are either slow at inference time (autoregressive models) or cannot generate variable-length, long-form audio (the rest).
Our work relies on latent diffusion to generate long-form (up to 95 sec), variable-length (controlled by the timing condition) stereo signals at 44.1kHz in 8 sec on an A100 GPU (40GB VRAM).

\textbf{Timing conditioning} ---
The use of learned embeddings to condition music generation models on timing information was introduced by Jukebox \cite{dhariwal2020jukebox}, an autoregressive model conditioned with timing information on: (i)~song duration, (ii)~starting time of the training/generated audio sample within the song, and (iii)~how much fraction of the song has elapsed.
We are not aware of previous works using timing conditioning for conditioning (latent) diffusion models.
Our work employs timing conditioning to control the length of the generations, enabling our latent diffusion models to generate variable-length outputs.

\textbf{Evaluation metrics} --- The commonly used quantitative audio metrics were developed for evaluating short-form mono audio generations at 16kHz~\cite{kilgour2018fr,copet2023simple}. Yet, our work focuses on generating long-form full-band stereo signals. Only \citet{pascual2023full} explored quantitative metrics for evaluating full-band audio, although their focus was short-form mono signals. Our work explores new quantitative metrics to evaluate long-form full-band stereo generations.
Qualitative metrics assessing audio quality and text alignment are also prevalent in the literature \cite{dong2023clipsonic,copet2023simple,ziv2024masked}. Our work also explores additional qualitative metrics to evaluate musicality, stereo correctness, and musical structure.

\textbf{Multitask generative modeling} --- 
While generative models have traditionally focused on specific tasks like speech, music or sound synthesis, recent works showed success in addressing all these tasks simultaneously \cite{yang2023uniaudio,liu2023audioldm2}. 
Our work relies on one model to generate both music and sound (not speech) given a text prompt.

\section{Architecture}
\label{architecture}

\begin{figure*}[!t]
  \centering
      \vspace{-1mm}
  \includegraphics[width=0.9\textwidth]{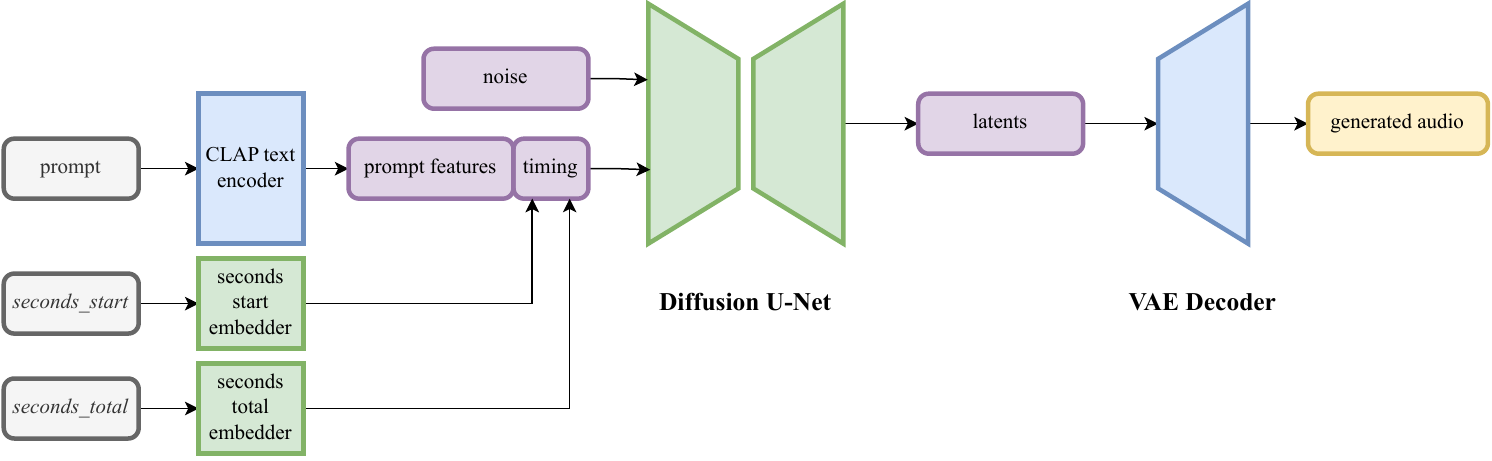}
  \vspace{-2mm}
  \caption{\textit{Stable Audio.} Blue: frozen pre-trained models. Green: parameters learnt during diffusion training. Purple: signals of interest.}
  \label{fig:stableaudio}
    \vspace{-1mm}
\end{figure*}
\begin{figure*}[!t]
  \centering
  \includegraphics[width=0.75\textwidth]{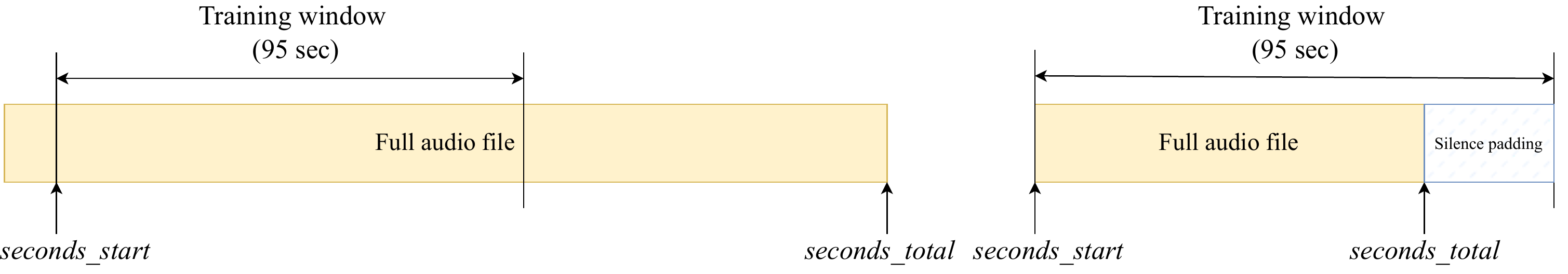}
  \vspace{-3mm}
  \caption{\textit{Timing embeddings examples.} \textit{Left}: Audio file longer than training window. \textit{Right}: Audio file shorter than training window.}
  \label{fig:timing}
    \vspace{-2mm}
\end{figure*}

Stable Audio is based on a latent diffusion model consisting of a variational autoencoder~(Section~\ref{vae}), a conditioning signal~(Section~\ref{conditioning}), and a diffusion model~(Section~\ref{diffusionmodel}).

\subsection{Variational autoencoder (VAE)}
\label{vae}

The VAE \cite{kingma2013autoencoding} compresses 44.1kHz stereo audio into an 
invertible (lossy) latent encoding that enables faster generation and training time compared to working with raw audio samples. To allow for arbitrary-length audio encoding and decoding, we use a fully-convolutional architecture (133M parameters) that follows the Descript Audio Codec \cite{kumar2023highfidelity} encoder and decoder (without the quantizer). 
We found that the Snake activations \cite{ziyin2020neural} in the Descript Audio Codec architecture improved 
audio reconstruction at high compression ratios compared to alternatives such as EnCodec \cite{défossez2022high}, at the expense of increased VRAM consumption. 
The VAE is trained from scratch on our dataset and downsamples the input stereo audio sequence by a factor of 1024, with the resulting latent sequence having a channel dimension of 64 (i.e., maps a $2 \times L$ input into $64 \times L/1024$ latent). This results in an overall data compression ratio of 32.

\subsection{Conditioning}
\label{conditioning}

\textbf{Text encoder} ---
\label{textencoder}
To condition on text prompts, we use a CLAP text encoder trained from scratch on our dataset. 
We use the actual setup recommended by the CLAP authors: (i) a HTSAT-based audio encoder with fusion having 31M parameters, and (ii) a RoBERTa-based text encoder of 110M parameters, both trained with a language-audio contrastive loss.
We use CLAP embeddings (instead of the also commonly used T5 embeddings) because its multimodal nature (language-audio) allows the text features to contain some information about the relationships between words and audio. 
Further, in Section \ref{ablation} we empirically note that the CLAP embeddings trained from scratch on our dataset can outperform the open-source CLAP and T5 embeddings. 
As shown by \citet{NovelAI_2022} when using CLIP \cite{radford2021learning} text features for Stable Diffusion \cite{rombach2022highresolution}, the text features in the next-to-last layer of the text encoder can provide a better conditioning signal than the text features from the final layer. Because of this, we use the text features from the next-to-last hidden layer of the CLAP text encoder. These text features are provided to the diffusion U-Net through cross-attention layers.

\textbf{Timing embeddings} ---
\label{timingembeds}
We calculate two properties 
when gathering a chunk of audio from our training data: the second from which the chunk starts (termed \textit{seconds\_start}) and the overall number of seconds in the original audio file (termed \textit{seconds\_total}), see Figure \ref{fig:timing}. For example, if we take a 95 sec chunk from an 180 sec audio file with the chunk starting at 14 sec, then \textit{seconds\_start} is 14 and \textit{seconds\_total} is 180 (see Figure \ref{fig:timing}, Left). These 
values are then translated into per-second discrete learned embeddings\footnote{We have a learnt, continuous timing embedding per second.} and concatenated along the sequence dimension with the text features from the prompt conditioning before being passed into the U-Net’s cross-attention layers. 
For training with audio files shorter than the training window (see Figure \ref{fig:timing}, Right), we pad with silence up to the training window length.
During inference, \textit{seconds\_start} and \textit{seconds\_total} are also provided 
as conditioning, allowing the user to specify the overall length of the output audio. For example, given our 95 sec model, setting \textit{seconds\_start} to 0 and \textit{seconds\_total} to 30 will create an output with 30 sec of audio followed by 65 sec of silence. 
This method allows the user generating variable-length music and sound effects. 

\subsection{Diffusion model}
\label{diffusionmodel}
Based on a U-Net (907M parameters) inspired by Mo\^{u}sai's architecture \cite{schneider2023mousai},  
it consists of 4 levels of symmetrical downsampling encoder blocks and upsampling decoder blocks, with skip connections between the encoder and decoder blocks providing a residual path at the same resolution. The 4 levels have channel counts of 1024, 1024, 1024, and 1280, and downsample by factors of 1 (no downsampling), 2, 2, and 4 respectively. After the final encoder block, there is a 1280-channel bottleneck block. 
Each block consists of 2 convolutional residual layers followed by a series of self-attention and cross-attention layers. Each encoder or decoder block has three of these attention layers, except for those in the first U-Net level, which only have one.
We rely on a fast and memory-efficient attention implementation \cite{dao2022flashattention}, to allow the model to scale more efficiently to longer sequence lengths. The diffusion timestep conditioning is passed in through FiLM layers \cite{perez2017film} to modulate the model activations based on the noise level. The prompt and timing conditioning information is passed in to the model through cross-attention layers. Further implementation details are in Appendix \ref{implementation}.

\subsection{Inference}
\label{inference}

Our sampling strategy during inference is based on the DPM-Solver++ \cite{lu2022dpm}, and we use classifier-free guidance (with a scale of 6) as proposed by \citet{lin2024common}. 
We use 100 diffusion steps during inference, see Appendix \ref{sec:steps} to know more on how the number of steps was chosen. 
Stable Audio is designed for variable-length, long-form music and sound generation. This is achieved by generating content within a specified window length (95 sec), and relying on the timing condition to fill the signal up to the length specified by the user and fill the rest with silence. 
To present variable-length audios (shorter than window length) to the end user, one can simply trim the silence. 
In Section \ref{sec:timing} we note that the timing conditioning is very reliable, showing the robustness of the proposed silence-trimming strategy.

\section{Training}
\label{training}

\subsection{Dataset}
\label{dataset}

Our dataset consists of 806,284 audios (19,500 hours) containing music (66\% or 94\%)\footnote{\label{note1}Percentages: number of files or GBs of content, respectively.}, sound effects (25\% or 5\%)\footnoteref{note1}, and instrument stems (9\% or 1\%)\footnoteref{note1}, with the corresponding text metadata from the stock music provider AudioSparx. Our dataset (audio and metadata) is online 
\footnote{\href{https://www.audiosparx.com}{https://www.audiosparx.com}} for consultation.

\subsection{Variational autoencoder (VAE)}
\label{vaetraining}

It was trained using automatic mixed precision for 1.1M steps 
with an effective batch size of 256 on 16 A100 GPUs. 
After 460,000 steps the encoder was frozen and the decoder was fine-tuned for an additional 640,000 steps. 
To ensure a consistent stereo reconstruction, we use a multi-resolution sum and difference STFT loss designed for stereo signals \cite{steinmetz2020automatic}. 
To that end, we apply A-weighting \cite{10.1121/1.1915637} before the STFT and use window lengths of 2048, 1024, 512, 256, 128, 64, and 32.
We also employ adversarial and feature matching losses using a multi-scale STFT discriminator modified to accept stereo audio \cite{défossez2022high}. The discriminators (set with 2048, 1024, 512, 256, and 128 STFT window lengths) use a complex STFT representation of the real and reconstructed audio, and a patch-based discriminative objective using the hinge loss \cite{défossez2022high}. 
Each loss is weighted as follows: 1.0 for spectral losses, 0.1 for adversarial losses, 5.0 for the feature matching loss, and 1e-4 for the KL loss.

\subsection{Text encoder}
\label{claptraining}
The CLAP model was trained for 100 epochs on our dataset from scratch, with an effective batch size of 6,144 with 64 A100 GPUs. We use the setup recommended by CLAP authors and train it with a language-audio contrastive loss. 

\subsection{Diffusion model}
\label{diffusiontraining}
It was trained using exponential moving average and automatic mixed precision for 640,000 steps on 64 A100 GPUs with an effective batch size of 256. 
The audio was resampled to 44.1kHz and sliced to 4,194,304 samples (95.1 sec). Files longer than this length were cropped from a random starting point, while shorter files were padded at the end with silence. We implemented a v-objective \cite{salimans2022progressive} with a cosine noise schedule and continuous denoising timesteps. We apply dropout (10\%) to the conditioning signals to be able to use classifier-free guidance. The text encoder is frozen while training the diffusion model.

\subsection{Prompt preparation}
\label{promptgeneration}
Each audio file in our dataset is accompanied by text metadata describing the audio file. This text metadata includes natural-language descriptions of the audio file's contents, as well as domain-specific metadata such as BPM, genre, moods, and instruments for music tracks. 
During the training of the text encoder and the diffusion model, we generate text prompts from this metadata by concatenating a random subset of the metadata as a string.
This allows for specific properties to be specified during inference, while not requiring these properties to be present at all times.
For half of the samples, we include the metadata-type (e.g., \textit{Instruments} or \textit{Moods}) and join them with the | character (e.g. \textit{Instruments: Guitar, Drums, Bass Guitar|Moods: Uplifting, Energetic}). For the other half, we do not include the metadata-type and join the properties with a comma (e.g. \textit{Guitar, Drums, Bass Guitar, Uplifting, Energetic}). For metadata-types with a list of values, we shuffle the list.

\section{Methodology}
\label{methodology}

\subsection{Quantitative metrics}
\label{quantitative}

\textbf{$\text{FD}_{openl3}$} --- The Fréchet Distance (FD) is utilized to evaluate the similarity between the statistics of a generated audio set and a reference audio set in a feature space. A low Fréchet Distance implies that the generated audio is plausible and closely matches the reference audio \cite{kilgour2018fr,copet2023simple}. While most previous works project the audio into the VGGish feature space \cite{hershey2017cnn}, we propose projecting it into the Openl3\footnote{The Openl3 settings we use: mel256 input, 44.1kHz, `music' or `env' content type depending if we evaluate music or audio, embedding size of 512, and hop size of 0.5 sec.} feature space \cite{cramer2019look}. Importantly, Openl3 accepts signals of up to 48kHz while VGGish operates at 16kHz.
    With this modification, our FD is not limited to evaluate downsampled 16kHz audio but it can evaluate the full bandwidth of the generated audios.
    Since we focus on generating 44.1kHz audio, we resample all the evaluation audios to 44.1kHz. Finally, we also extend the FD to evaluate stereo signals. To that end, we project left- and right-channel audios into Openl3 features independently, and concatenate them to obtain the stereo features. If the evaluation audio is mono, we concatenate copied Openl3 (mono) features to obtain the desired stereo features. Hence, we propose a novel $\text{FD}_{openl3}$ metric to study the plausibility of the generated variable-length, full-band stereo signals.
    
 \textbf{$\text{KL}_{passt}$} --- We use PaSST, a state-of-the-art audio tagger trained on AudioSet \cite{koutini22passt}, to compute the Kullback–Leibler (KL) divergence over the probabilities of the labels between the generated and the reference audio \cite{copet2023simple}. The generated audio is expected to share similar semantics (tags) with the reference audio when the KL is low. While most previous works focus on generating short snippets, our work focuses on generating long-form audio. For this reason, we modify the KL to evaluate audios of varying and longer lengths. This adaptation involves segmenting the audio into overlapping analysis windows\footnote{PaSST model was originally trained with 10 sec inputs, and we utilize an analysis window of 10 sec (to match PaSST training) with a 5 sec overlap (50\% overlap, for compute efficiency).}. Subsequently, we calculate the mean (across windows) of the generated logits and then apply a softmax. Finally, PaSST operates at 32kHz. To evaluate our 44.1kHz models, we resample all the evaluation audios from 44.1kHz to 32kHz. Hence, we propose a novel $\text{KL}_{passt}$ metric capable to evaluate the semantic correspondence between lengthy generated and reference audios up to 32kHz.
    
 \textbf{$\text{CLAP}_{score}$} ---  The cosine similarity is computed between the $\text{CLAP}_{\text{LAION}}$ text embedding of the given text prompt and the $\text{CLAP}_{\text{LAION}}$ audio embedding of the generated audio \cite{wu2023large,huang2023make}. A high $\text{CLAP}_{score}$ denotes that the generated audio adheres to the given text prompt.
    Differently from previous works, that evaluate 10 sec inputs, we use the `feature fusion' variant of $\text{CLAP}_{\text{LAION}}$ to handle longer audios. It is based on `fusing' (concatenating) inputs at various time-scales: a global input (downsampled to be of 10 sec) is concatenated to 3 random crops (of 10 sec) from the first, middle, and last parts of the audio.
    $\text{CLAP}_{\text{LAION}}$ audio embeddings are computed from 48kHz audio. To evaluate our 44.1kHz models, we resample all the evaluation audios from 44.1kHz to 48kHz.
    Hence, we propose a novel $\text{CLAP}_{score}$ to evaluate how 48kHz audios longer than 10 sec adhere to a given text prompt. %$\text{CLAP}_{score}$ also evaluates audios at 48kHz.
    \footnotetext{Used checkpoint: `630k-audioset-fusion-best'.} 

In short, we adapted established metrics to assess the more realistic use case of long-form full-band stereo generations. All quantitative metrics can deal with variable-length inputs.

\subsection{Qualitative metrics} 
\label{qualitative}

\textbf{Audio quality} --- We evaluate whether the generated audio is of low-fidelity with artifacts or high-fidelity.

\textbf{Text alignment} --- We evaluate how the generated audio adheres to the given text prompt.

\textbf{Musicality} (music only) --- We evaluate the capacity of the model to articulate melodies and harmonies.

\textbf{Stereo correctness} (stereo only) --- We evaluate the appropriateness of the generated spatial image.

\textbf{Musical structure} (music only) --- We evaluate if the generated song contains intro, development, and/or outro.

We collect human ratings for the  metrics above and report mean opinion scores for audio quality, text alignment, and musicality in the following scale: bad (0), poor (1), fair (2), good (3) and excellent (4).
We observed that assessing stereo correctness posed a significant challenge for many users. To address this, we streamlined the evaluation by seeking for a binary response: either stereo correctness or not. Similarly, we adopted a binary approach for evaluating musical structure. We ask users to determine whether the generated music exhibits some common structural elements of music (intro, development, outro) or not.
For those binary responses (stereo correctness and musical structure) we report percentages.
Note that musicality and musical structure are only evaluated for music signals. For non-music (audio) signals we evaluate audio quality, text alignment and stereo correctness.
Also note that stereo correctness is only evaluated for stereo signals. 
We relied on webMUSHRA \cite{schoeffler2018webmushra} to run our perceptual experiments. 
We are not aware of previous works that qualitatively assess musicality, stereo correctness, and/or musical structure.

\subsection{Evaluation data}

\textbf{Quantitative experiments} --- We rely on the standard MusicCaps \cite{agostinelli2023musiclm} and AudioCaps \cite{kim2019audiocaps} benchmarks. 
MusicCaps contains 5,521 music segments from YouTube, each with 1 caption (5,434 audios were available for download).
AudioCaps test set contains 979 audio segments from YouTube, each with several captions (881 audios were available for download, and it includes 4,875 captions).
For every model to evaluate, we generate an audio per caption. This results in 5,521 generations for the MusicCaps evaluations and 4,875 generations for the AudioCaps ones.
While these benchmarks are not typically used for evaluating full-band stereo signals, the original data is predominantly stereo and full-band (Appendix \ref{sec:original}). We rely on the original data resampled to 44.1kHz to meet the target bandwidth of Stable Audio. 
Finally, since the standard MusicCaps and AudioCaps segments are of 10 sec, we also looked into the full-length audios to consider variable-length long-form evaluation content. Yet, captions do not hold consistently throughout the whole (long) audio, as they only accurately represent the intended 10 sec segment. As a result, reference audios are of 10 sec while generated audios range from 10 to 95 sec (Tables \ref{tab:musiccaps} and \ref{tab:audiocaps}).
Hence, in addition to modifying the established metrics to evaluate full-band stereo generations, it was also crucial to adapt the standard datasets to align with our evaluation criteria.

\textbf{Qualitative experiments} --- Prompts for qualitative evaluation were randomly picked from MusicCaps and AudioCaps. We avoided prompts including "low quality" (or similar) to focus on high-fidelity synthesis, avoided ambient music because users found challenging to evaluate musicality, and avoided speech-related prompts since it is not our focus.

\begin{table*}[]
\centering
\begin{tabular}{lccccccc}
\toprule
                          &  & output  & & & & inference  \\
                          & channels/sr &  length & $\text{FD}_{openl3}$ $\downarrow$ & $\text{KL}_{passt}$ $\downarrow$ & $\text{CLAP}_{score}$ $\uparrow$ &  time \\ \midrule
Training data (upper bound) & 2/44.1kHz & full songs &  101.47  & -   &     - & - \\
Autoencoded training data & 2/44.1kHz & full songs &  117.52  & -   &     - & - \\ 
\midrule
%Stable Audio w/ T5   & 2/44.1kHz & 95 sec &  115.53 & 1.04 &  0.41   & ? sec \\ 
Stable Audio w/ $\text{CLAP}_{\text{ours}}$   & 2/44.1kHz & 23 sec & \underline{118.09} & \underline{0.97} &  \underline{0.44}  & 4 sec \\ 
Stable Audio w/ $\text{CLAP}_{\text{LAION}}$    & 2/44.1kHz & 23 sec & 123.30 & 1.09 &  0.43  & 4 sec \\  
Stable Audio w/ T5   & 2/44.1kHz & 23 sec & 126.93 & 1.06 & 0.41  & 4 sec \\ 
\midrule
AudioLDM2-music  & 1/16kHz & 95 sec &  354.05 &  1.53 &  0.30   & 38 sec \\
{AudioLDM2-large}  & 1/16kHz & 95 sec &  339.25 &  {1.46} &  0.30   & 37 sec \\           
{AudioLDM2-48kHz } & 1/\textbf{48kHz} & 95 sec &  {299.47}  &  {2.77}  &  {0.22}    & 242 sec \\  
%MusicGen-small (audiosparx)  & 1/32kHz &95 sec & 216.86  &  1.17 &  0.29   & 128 sec \\
MusicGen-small   & 1/32kHz & 95 sec &  205.65 & 0.96  &  0.33   & 126 sec \\
{MusicGen-large}   & 1/32kHz  & 95 sec &  197.12 &  {0.85} &   0.36  & 242 sec \\
MusicGen-large-stereo   & \textbf{2}/32kHz  & 95 sec &  216.07 &  1.04 &   0.32  & 295 sec \\  % /fsx/jordipons/musicgenstereo (running) -> 5521 files. 
{Stable Audio} & \textbf{2}/{44.1kHz} & {95 sec} &  \textbf{108.69} & \textbf{0.80}  &   \textbf{0.46}  & \textbf{8} sec \\ %10 sec
%Audiosparx (upper bound) & 2/44.1kHz & full songs/audio &  98.72  & -   &     - & - \\
\bottomrule
\end{tabular}
\caption{\textit{Quantitative results on MusicCaps}. Top: autoencoder audio fidelity study, discussed in Section \ref{sec:autoencoder-quality}. Middle: text encoder ablation study, discussed in Section \ref{ablation}. Bottom: comparing Stable Audio against the state-of-the-art, see Section \ref{sec:sota-comparison}. Different experiments (top, middle, bottom sections of the table) are not strictly comparable due to different output lengths. \underline{Underlines} denote the best results in the middle section of the table, and \textbf{bold} indicates the best results in the bottom section.}
\label{tab:musiccaps}
\end{table*}

\begin{table*}[]
\centering
\begin{tabular}{lccccccc}
\toprule
                          &  & output  & & & & inference  \\
                          & channels/sr &  length & $\text{FD}_{openl3}$ $\downarrow$ & $\text{KL}_{passt}$ $\downarrow$ & $\text{CLAP}_{score}$ $\uparrow$ &  time \\ \midrule
Training data (upper bound) & 2/44.1kHz & full-length audio  &  88.78  & -   &     - & - \\  
Autoencoded training data & 2/44.1kHz & full-length audio &  106.13  & -   &     - & - \\ \midrule 
Stable Audio w/ $\text{CLAP}_{\text{ours}}$   & 2/44.1kHz & 23 sec & \underline{114.25} & \underline{2.57} &  0.16  & 4 sec \\ 
Stable Audio w/ $\text{CLAP}_{\text{LAION}}$    & 2/44.1kHz & 23 sec & 119.29 & 2.73 &  \underline{0.19}  & 4 sec \\  
Stable Audio w/ T5   & 2/44.1kHz & 23 sec & 119.28 & 2.69 &  0.11  & 4 sec \\ \midrule                 
{AudioLDM2-large}  & 1/16kHz & 10 sec &  170.31  &  1.57 &  0.41   & 14 sec \\  
AudioLDM2-48kHz  & 1/\textbf{48kHz} & 10 sec &  \textbf{101.11} &  2.04  &  0.37    & 107 sec \\ 
{AudioGen-medium}   & 1/16kHz  & 10 sec &  186.53 &  \textbf{1.42} &  \textbf{0.45}  & 36 sec \\ 
{Stable Audio}    & \textbf{2}/44.1kHz & 95 sec $^\dagger$ &  \textbf{103.66} & 2.89 &  0.24 & \textbf{8} sec \\

\bottomrule
\end{tabular}
\caption{\textit{Quantitative results on AudioCaps}. Top: autoencoder audio fidelity study, discussed in Section \ref{sec:autoencoder-quality}. Middle: text encoder ablation study, discussed in Section \ref{ablation}. Bottom: comparing Stable Audio against the state-of-the-art, see Section \ref{sec:sota-comparison}. Different experiments (top, middle, bottom sections of the table) are not strictly comparable due to different output lengths. \textbf{$^\dagger$} Stable Audio was trained to generate 95 sec outputs, but during inference it can generate variable-length outputs by relying on the timing conditioning. Despite Stable Audio generating 95 sec outputs and the rest of state-of-the-art models generating 10 sec outputs, it is still significantly faster. We trim audios to 10 sec (discarding the end silent part) for a fair quantitative evaluation against the state-of-the-art (see Section \ref{inference} for inference details).}

\label{tab:audiocaps}
\end{table*}

\begin{table*}[]
\centering
\begin{tabular}{lcccc|ccc}
\toprule
& \multicolumn{4}{c|}{MusicCaps} & \multicolumn{3}{c}{AudioCaps} \\ \midrule
& Stable & \multicolumn{1}{|c|}{MusicGen} & \multicolumn{1}{|c|}{MusicGen} & \multicolumn{1}{|c|}{AudioLDM2} & Stable & \multicolumn{1}{|c|}{AudioGen} & \multicolumn{1}{|c}{AudioLDM2} \\
& Audio  & \multicolumn{1}{|c|}{large}  & \multicolumn{1}{c|}{stereo} &  48kHz  & Audio  & \multicolumn{1}{|c|}{medium}  &  48kHz \\ \midrule      
Audio Quality & \textbf{3.0}$\pm$0.7 & 2.1$\pm$0.9 & 2.8$\pm$0.7 & 1.2$\pm$0.5 & \textbf{2.5}$\pm$0.8 & 1.3$\pm$0.4 & 2.2$\pm$0.9 \\  
Text Alignment & \textbf{2.9}$\pm$0.8 & 2.4$\pm$0.9 & 2.4$\pm$0.9 & 1.3$\pm$0.6 & 2.7$\pm$0.9 & 2.5$\pm$0.9 & \textbf{2.9}$\pm$0.8 \\ 
Musicality  & \textbf{2.7}$\pm$0.9 & 2.0$\pm$0.9 & \textbf{2.7}$\pm$0.9 & 1.5$\pm$0.7 &       -             &       -    & -\\ \midrule
Stereo correctness & \textbf{94.7}\% &     -      & 86.8\% &   -      & 57\% &     -      &  -\\
Structure: intro & \textbf{92.1}\% & 36.8\%&52.6\%& 2.6\% &       -             &      -     & - \\
Structure: development & 65.7\%&68.4\%&\textbf{76.3}\%&15.7\%&       -            &     -    &   - \\
Structure: outro& \textbf{89.4}\% & 26.3\% & 15.7\% &2.6\%&       -            &      -     &-\\
\bottomrule
\end{tabular}
\caption{\textit{Qualitative results}. Top: mean opinion score $\pm$ standard deviation. Bottom: percentages. 19 users participated in this study.}
\label{tab:perceptual}
\end{table*}

\subsection{Baselines}
\label{baselines}

Direct comparisons with some models (e.g., Mo\^{u}sai or JEN1) is infeasible {as their weights are not accessible.} For this reason, we benchmark against AudioLDM2, MusicGen, and AudioGen. 
These are state-of-the-art open-source models representative of the current literature: latent diffusion models (AudioLDM2) or autoregressive models (MusicGen, AudioGen), that can be stereo (MusicGen-stereo) or mono, and at various sampling rates (see Table \ref{tab:musiccaps} and \ref{tab:audiocaps}).
The AudioLDM2 variants we evaluate are: `{AudioLDM2-48kHz}' that was trained to generate full-band mono sounds and music,
`{AudioLDM2-large}' to generate 16kHz mono sounds and music,
and `{AudioLDM2-music}' that was trained on music only to generate 16kHz mono music (checkpoints\footnote{The used checkpoints are `audioldm\_48k', `audioldm2-full-large-1150k' and `audioldm2-music-665k', respectively.}). 
The MusicGen variants we evaluate are: `MusicGen-small' that is a compute-efficient autoregressive model for music generation, `MusicGen-large' that is its large variant, and `MusicGen-large-stereo' that is its stereo version. 
However, MusicCaps includes vocal-related prompts and MusicGen models are not trained to generate vocals. In Appendix \ref{nosinging} we also benchmark against MusicGen without vocal prompts.
We also evaluate `AudioGen-medium', the only open-source {autoregressive model available for sound synthesis.}

\section{Experiments}

\subsection{How does our autoencoder impact audio fidelity?}
\label{sec:autoencoder-quality}

To understand the reconstruction capabilities of our latent space, we project a subset of training data {(5,521 and} 4,875 audios in Table \ref{tab:musiccaps} and \ref{tab:audiocaps}, respectively) through our autoencoder to obtain the latents and reconstruct from them. Then, we compare the $\text{FD}_{openl3}$ of the real and the autoencoded training data with respect to the MusicCaps and AudioCaps evaluation audio (Tables \ref{tab:musiccaps} and \ref{tab:audiocaps}). 
In both cases, the autoencoded training data yields slightly inferior results compared to the real training data. This indicates a marginal degradation, yet informal listening suggests that the impact is fairly transparent (examples available in our demo website).

\subsection{Which text encoder performs the best?}
\label{ablation}

Various text encoders are prevalent in the literature, including: the open-source CLAP \cite{wu2023large} denoted here as $\text{CLAP}_{\text{LAION}}$, privately trained CLAP-like models denoted here as $\text{CLAP}_{\text{ours}}$ (trained as in Section \ref{claptraining}), and the open-source T5 embeddings. An ablation study is conducted in Tables \ref{tab:musiccaps} and \ref{tab:audiocaps} to determine which text encoder performs the best. In this study, we train the base diffusion model in Section \ref{diffusiontraining} for 350k steps with different text encoders and evaluate them using our qualitative metrics both on Musiccaps and AudiocCaps. The text encoders are frozen during training. Results indicate comparable performance, with $\text{CLAP}_{\text{ours}}$ exhibiting a slight superiority, leading us to choose it for further experimentation. The utilization of a privately trained CLAP guarantees the use of text embeddings trained on the same dataset as our diffusion model. This approach ensures consistency across all components of the model, mitigating distribution or vocabulary mismatches between the text embeddings and the diffusion model.

\subsection{How accurate is the timing conditioning?}
\label{sec:timing}

\begin{figure}[t]
\begin{center}
\centerline{\includegraphics[width=0.9\columnwidth]{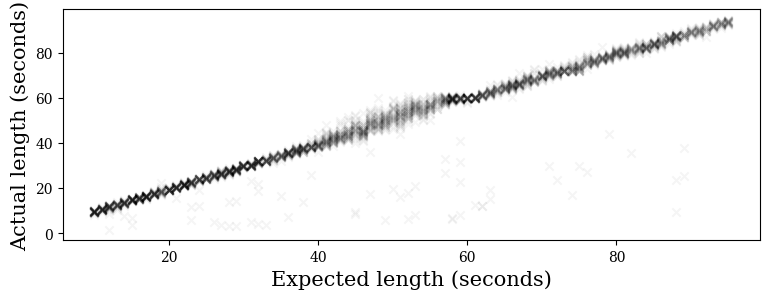}}
\vspace{-4mm}
\caption{Comparing the actual length (measured in the signal) against the expected length (provided by the timing conditioning).}
\vspace{-9mm}
\label{fig:timingeval}
\end{center}
\end{figure}

The timing condition is evaluated by generating audios of variable lengths (length controlled by the timing condition) to note its behavior across different length values \mbox{(Figure \ref{fig:timingeval}).}
We compare the expected length (provided by the timing conditioning) against the measured one, aiming for a diagonal in Figure \ref{fig:timingeval}. We measure the length of the audio by detecting when the signal becomes silence with a simple energy threshold---because, e.g., a model with a 30 sec timing condition is expected to fill the 95 sec window with 30 sec of signal plus 65 sec of silence.
In Figure \ref{fig:timingeval} we note that the model is consistently generating audios of the expected length, with more errors around 40-60 sec. This error might be caused because there is less training data of this duration.
Also, note that some of the shortest measured lengths (seen in gray) may be false positives resulting from the simplistic silence detector we use.
Appendix \ref{app:timingcond} includes more results.

\subsection{How does it compare with the state-of-the-art?}
\label{sec:sota-comparison}

This section discusses Tables \ref{tab:musiccaps}, \ref{tab:audiocaps}, and \ref{tab:perceptual}.
Stable Audio can outperform the state-of-the-art in audio quality and also improves text alignment in MusicCaps. Yet, text alignment is slightly worse in AudioCaps possibly due to the small amount of sound effects in our training set (Section \ref{dataset}).
It is also very competitive at musicality and at generating correct stereo music signals. It's interesting, though, its low stereo correctness score in AudioCaps. It might be caused because the randomly selected prompts did not require much stereo movement, resulting in renders that are relatively non-spatial (see in our \href{https://stability-ai.github.io/stable-audio-demo/}{demo website}). 
Despite this difficulty, the stereo render remained consistent without artefacts, leading to a stereo correctness score of 57\%. Our \href{https://stability-ai.github.io/stable-audio-demo/}{demo website} includes more stereo sound examples.
Finally, Stable Audio is also capable to generate structured music: with intro, some degree of development, and outro.
Note that state-of-the-art models are not consistent at generating a coherent structure, since they are mainly capable of developing musical ideas.

\subsection{How fast is it?}
\label{fast}

We compare inference times using one A100 GPU and a batch size of 1. First, note that latent diffusion (AudioLDM2 and Stable Audio) is significantly faster than autoregressive modeling, as outlined in the introduction. Second, note that Stable Audio (operating at stereo 44.1kHz) is also faster than AudioLDM2-large and -music (operating at mono 16kHz). Stable Audio's speedup is even more significant when compared to AudioLDM2-48kHz (operating at mono 48kHz)\footnote{AudioLDM2-large and -music are implemented with Diffusers, 3x faster than the native implementation of the 48kHz one. AudioLDM2 runs use the setup recommended by the authors.}.

\section{Conclusions}
\label{conclusions}

Our latent diffusion model enables the rapid generation of variable-length, long-form stereo music and sounds at 44.1kHz from textual and timing inputs.
We explored novel qualitative and quantitative metrics for evaluating long-form full-band stereo signals, and found Stable Audio to be a top contender, if not the top performer, in two public benchmarks.
Differently from other state-of-the-art models, ours can generate music with structure and stereo sound effects.

\section*{Acknowledgments}

Thanks to J. Parker and Z. Zukowski for their feedback, and to the qualitative study participants for their contributions.

\section*{Impact Statement}

Our technology represents a significant improvement in assisting humans with audio production tasks, offering the capability to generate variable-length, long-form stereo music and sound effects based on text descriptions. This innovation expands the toolbox available to artists and content creators, enriching their creativity. However, alongside its potential benefits, also confronts several inherent risks.
One prominent concern lies in the reflection of biases present in the training data. This raises questions about the appropriateness of the technology for cultures underrepresented in the training dataset. 
Moreover, the contextual nature embedded in audio recordings and music emphasize the importance of careful consideration and collaboration with stakeholders.
In light of these considerations, we commit to continued research and collaboration with stakeholders (like artists and data providers) to navigate the complex landscape of AI-based audio production responsibly.

\bibliography{example_paper}
\bibliographystyle{icml2023}

\newpage
\appendix
\onecolumn
\section{Inference diffusion steps}
\label{sec:steps}

\begin{figure}[ht]
\begin{center}
\vspace{-4mm}
\centerline{\includegraphics[width=0.50\columnwidth]{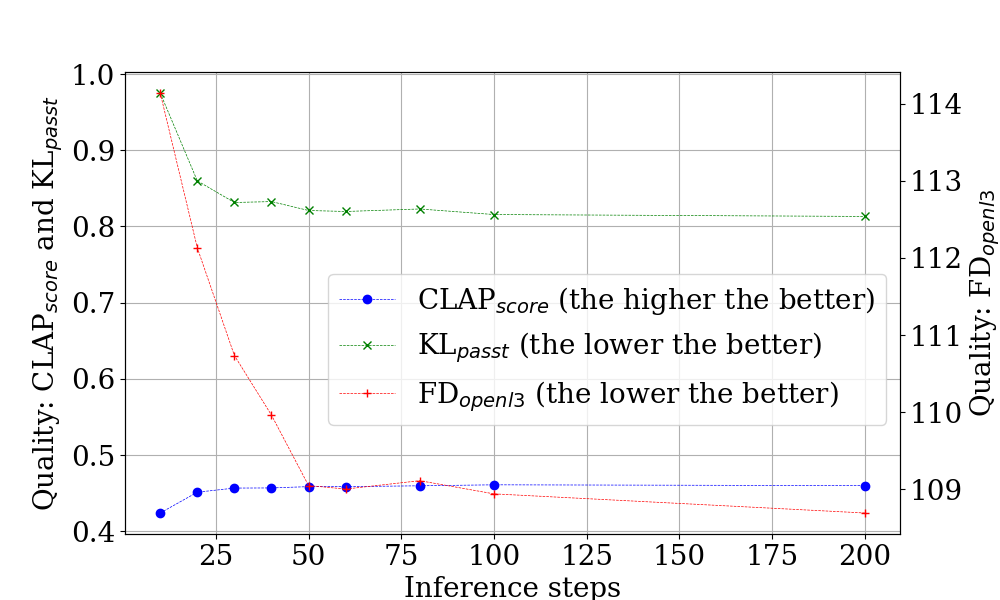}}
\vspace{-2mm}
\caption{Quality metrics vs Inference diffusion steps (trade-off).}
\label{steps}
\end{center}
\vspace{-4mm}
\end{figure}

In diffusion generative modeling, a critical consideration revolves around the trade-off between the quality of the generated outputs and the number of inference steps employed (quality vs inference-speed trade-off). Our results in Figure \ref{steps} show that a significant portion of the overall improvement in output quality is achieved within the initial 50 inference steps, suggesting diminishing returns with additional computational effort. Given that, we choose to set the total inference steps to 100. This decision is undertaken with a precautionary approach, ensuring a sufficient number of time steps to guarantee certain quality in the generated outputs.
Nevertheless, this implies the possibility of being more aggressive with the number of diffusion steps to significantly accelerate our inference times without  compromising much quality.

\section{MusicCaps and AudioCaps: the original data from Youtube}
\label{sec:original}

\begin{figure}[ht]
\begin{center}
\centerline{\includegraphics[width=0.50\columnwidth]{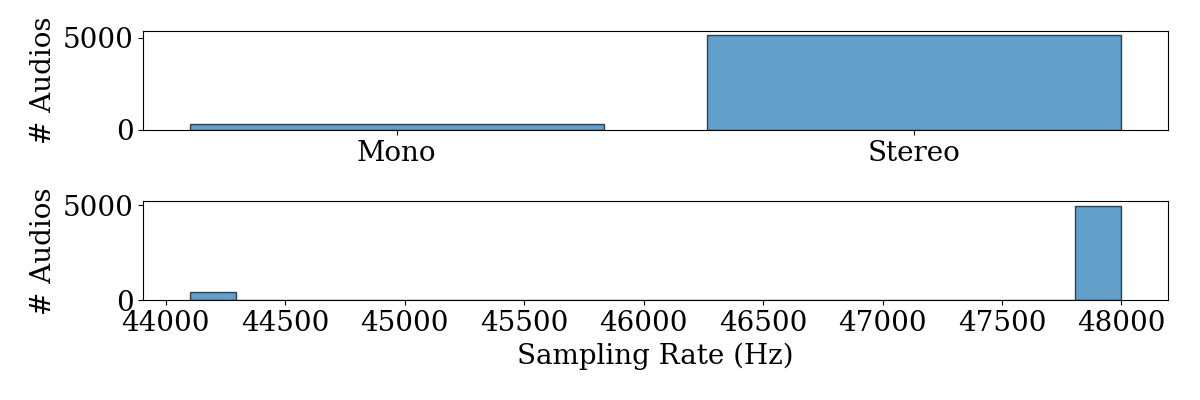}}
\vspace{-4mm}
\caption{Statistics of the MusicCaps original data.}
\label{fig:musiccaps}
\end{center}
\end{figure}

\begin{figure}[ht]
\begin{center}
\centerline{\includegraphics[width=0.50\columnwidth]{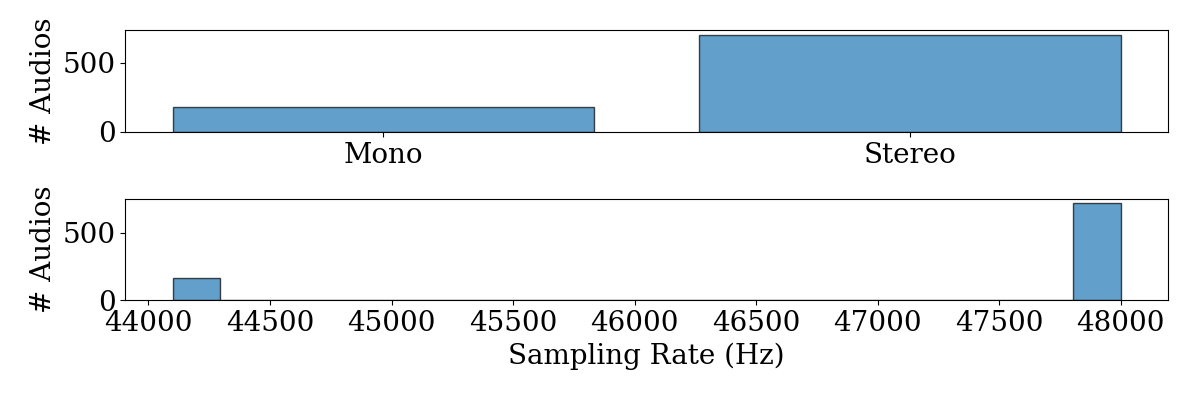}}
\vspace{-4mm}
\caption{Statistics of the AudioCaps original data.}
\label{fig:audiocaps}
\end{center}
\end{figure}

MusicCaps and AudioCaps benchmarks are not commonly used for evaluating full-band stereo signals, since most researchers typically use mono versions at 16kHz of those datasets.
However, the original data is predominantly stereo and full-band (see Figures \ref{fig:musiccaps} and \ref{fig:audiocaps}). 
Provided that this data is easily available, we rely on the original data resampled at 44.1kHz to meet the target bandwidth of Stable Audio.
This approach ensures that our evaluation encompasses the richness inherent in the original stereo full-band signals, providing a more accurate representation of the model's performance under conditions reflective of real-world data.


\section{Timing conditioning: additional evaluation}
\label{app:timingcond}

In Section \ref{sec:timing} we observed that Stable Audio adheres to the timing conditioning to generate signals of the specified length.

\begin{figure}[ht]
\begin{center}
\centerline{\includegraphics[width=0.50\columnwidth]{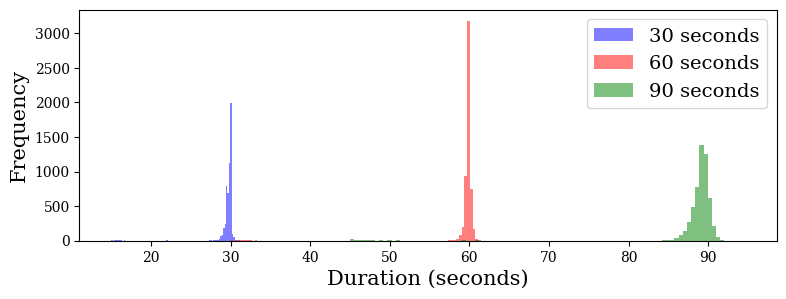}}
\vspace{-3mm}
\caption{Histogram depicting the measured lengths of MusicCaps captions.}
\label{fig:histogramtiming}
\end{center}
\end{figure}

We further study its behaviour by generating MusicCaps prompts at various lengths: 30, 60 and 90 sec. In Figure \ref{fig:histogramtiming} we depict the histogram of the measured lengths, clustered by the specified lengths (blue 30 sec, red 60 sec, and green 90 sec).
As in Section \ref{sec:timing}, we measure the length of the audio by detecting when the signal becomes silence with a simple energy threshold. %---because, e.g., a model with a 30 sec timing condition is expected to fill the 95 sec window with 30 sec of signal plus 65 sec of silence. 
In this experiment we note that the timing conditioning is fairly precise, generating audios of the expected length, with an error of a few seconds with a slight bias towards generating shorter audios. This means that the audio tends to finish right before the expected length, making it very appropriate to cut out the signal at the expected length.
Also, note that some of the shortest measured lengths may be attributed to false positives resulting from the simplistic silence detector we use.

\section{Related work: additional discussion on latent diffusion models}
\label{appendix:related}

Mo\^{u}sai and JEN-1 are closely related to our work. Both 
target high-fidelity stereo music synthesis with latent diffusion models. Our work, however, differs from Mo\^{u}sai in several key aspects:

\begin{itemize}
    \item Mo\^{u}sai's latent is based on a spectrogram-based encoder and a diffusion decoder that requires 100 decoding steps, while ours in a fully-convolutional end-to-end VAE. This distinction is crucial in achieving our fast inference times.
    \item Mo\^{u}sai's realtime factor is of $\times$1, while ours is of $\times$10.
    \item Mo\^{u}sai uses a form of timing conditioning based on information about chunked audio files in the prompts (e.g. \textit{Electro House, 3 of 4}), but we include explicit timing conditioning that allows for variable-length audio generation.
\end{itemize}

Our work differs from JEN-1 in the following aspects:

\begin{itemize}
    \item JEN-1 relies on a masked autoencoder with a dimensionality reduced latent, and also on a \textit{omnidirectional} latent diffusion model trained in a multitask fashion. In contrast, Stable Audio is inherently simpler with no \textit{omnidirectional} constraints, no dimensionality reduction, and no multitask training. Our approach allows for an easier implementation and training while still being able to deliver state-of-the-art results.
    \item JEN-1 is trained to generate 10 sec of music, not to generate variable-length, long-form music and sound effects.
\end{itemize}
Note that Mo\^{u}sai and JEN-1 target music synthesis while we target both music and sounds synthesis with a single model. 

AudioLDM2 is also closely related to our work. It is a latent diffusion model capable to generate mono speech, sound effects, and music up to 48kHz. Although the original AudioLDM2 was designed to operate at 16kHz, a recent release operates at 48kHz. Our work, however, differs from AudioLDM2 in several key aspects:

\begin{itemize}
    \item AudioLDM2 relies on a shared representation for music, audio, and speech to condition the latent diffusion model. This representation is shared between an audio masked auto encoder (audioMAE) and a GPT-2 that takes audio, text, speech transcripts, and images. As Stable Audio was not trained for speech generation or image-to-audio, there's no need to incorporate the intricacies of GPT-2. Instead, we opt for a CLAP text encoder.
    \item Stable Audio is faster and outperforms AudioLDM2 in audio quality and on text alignement for music generation. Yet, AudioLDM2 outperforms Stable Audio on text alignement for sound effects generation (see Tables \ref{tab:musiccaps}, \ref{tab:audiocaps}, and \ref{tab:perceptual}).
\end{itemize}

Mo\^{u}sai, JEN-1, and AudioLDM2 use the open-source, pretrained T5 or FLAN-T5 text embeddings, and we use CLAP text embeddings trained on the same dataset as our diffusion model. Our approach ensures consistency across all components of the model, eliminating distribution (or vocabulary) mismatches between the text embeddings and the diffusion model.

\section{Additional MusicCaps results: quantitative evaluation without singing-voice prompts}
\label{nosinging}

MusicCaps includes vocal-related prompts but MusicGen's released weights (used for benchmarking) are not trained to generate vocals\footnote{Vocals were removed from their training data source using the corresponding tags, and then using a music source separation method.}.
To allow for a fair evaluation against MusicGen, we also evaluate the models in Table \ref{tab:musiccaps} with a subset of 2184 prompts that do not include vocals\footnote{Prompts containing any of those words were removed: speech, speech synthesizer, hubbub, babble, singing, male, man, female, woman, child, kid, synthetic singing, choir, chant, mantra, rapping, humming, groan, grunt, vocal, vocalist, singer, voice, and acapella.}. In Table \ref{tab:nosinging}, we observe results akin to those in Table \ref{tab:musiccaps}: Stable Audio consistently obtains better results than the rest (with the exception of MusicGen-large that obtains comparable $\text{KL}_{passt}$ scores to ours).

\begin{table*}[h]
\centering
\begin{tabular}{lccccccc}
\toprule
                          &  & output  & & & & inference  \\
                          & channels/sr &  length & $\text{FD}_{openl3}$ $\downarrow$ & $\text{KL}_{passt}$ $\downarrow$ & $\text{CLAP}_{score}$ $\uparrow$ &  time \\ \midrule
AudioLDM2-music  & 1/16kHz & 95 sec &  354.37 &  1.66 &  0.32   & 38 sec \\
{AudioLDM2-large}  & 1/16kHz & 95 sec &  349.67 &  1.66 &  0.32   & 37 sec \\           
{AudioLDM2-48kHz } & 1/\textbf{48kHz} & 95 sec & 296.46 &  3.15 &  0.22   & 242 sec \\  
MusicGen-small   & 1/32kHz & 95 sec & 186.28 &  1.02 &  0.34  & 126 sec \\
{MusicGen-large}   & 1/32kHz  & 95 sec & 176.54 &  \textbf{0.86} &  0.37 & 242 sec \\
MusicGen-large-stereo   & \textbf{2}/32kHz  & 95 sec & 196.66 &  1.08 &  0.33 & 295 sec \\  
{Stable Audio} & \textbf{2}/{44.1kHz} & {95 sec} &  \textbf{100.29} &  \textbf{0.87} & \textbf{ 0.44}  & \textbf{8} sec \\ 
\bottomrule
\end{tabular}
\begin{center}
\vspace{-4mm}
\end{center}
\caption{\textit{Quantitative results on MusicCaps without singing-voice prompts}. Scores highlighted in \textbf{bold} indicate which are the best results.}
\label{tab:nosinging}
\end{table*}

\section{Implementation details}
\label{implementation}

Code to reproduce Stable Audio can be found online: \href{https://github.com/Stability-AI/stable-audio-tools}{https://github.com/Stability-AI/stable-audio-tools}. 

The configuration file used for training and defining our VAE autoencoder is \href{https://github.com/Stability-AI/stable-audio-tools/blob/main/stable_audio_tools/configs/model_configs/autoencoders/stable_audio_1_0_vae.json}{available online}. 

The configuration file used for training and defining our latent diffusion model is \href{https://github.com/Stability-AI/stable-audio-tools/blob/main/stable_audio_tools/configs/model_configs/txt2audio/stable_audio_1_0.json}{available online}. 

The provided configuration files offer compact descriptions of the architecture and implementation of Stable Audio. These configurations serve as additional resources for comprehensively understanding the underlying implementation of Stable Audio.

Code to reproduce our metrics can be found online: \href{https://github.com/Stability-AI/stable-audio-metrics}{https://github.com/Stability-AI/stable-audio-metrics}. 

We relied on the code shared by the CLAP authors \cite{wu2023large} to train our text encoder with our private dataset:
\vspace{-2mm}
\begin{center}
    \href{https://github.com/LAION-AI/CLAP}{https://github.com/LAION-AI/CLAP}
\end{center}

\end{document}